\documentclass[12 pt, amsfonts, amssymb,color]{article}

\usepackage{amsmath,amssymb,amsfonts,latexsym}
\usepackage[all]{xy}

\usepackage
{hyperref}
\begin{document}

\renewcommand\theequation{\arabic{section}.\arabic{equation}}
\catcode`@=11 \@addtoreset{equation}{section}
\newtheorem{defn}{Definition}[section]
\newtheorem{theorem}{Theorem}[section]
\newtheorem{axiom2}{Example}[section]
\newtheorem{lem}{Lemma}[section]
\newtheorem{prop}{Proposition}[section]
\newtheorem{cor}{Corollary}[section]
\newcommand{\be}{\begin{equation}}
\newcommand{\ee}{\end{equation}}

\def\forms{{\textstyle\bigwedge}}
\newcommand{\equal}{\!\!\!&=&\!\!\!}
\newcommand{\rd}{\partial}
\newcommand{\g}{\hat {\cal G}}
\newcommand{\bo}{\bigodot}
\newcommand{\res}{\mathop{\mbox{\rm res}}}
\newcommand{\diag}{\mathop{\mbox{\rm diag}}}
\newcommand{\Tr}{\mathop{\mbox{\rm Tr}}}
\newcommand{\const}{\mbox{\rm const.}\;}
\newcommand{\cA}{{\cal A}}
\newcommand{\bA}{{\bf A}}
\newcommand{\Abar}{{\bar{A}}}
\newcommand{\cAbar}{{\bar{\cA}}}
\newcommand{\bAbar}{{\bar{\bA}}}
\newcommand{\cB}{{\cal B}}
\newcommand{\bB}{{\bf B}}
\newcommand{\Bbar}{{\bar{B}}}
\newcommand{\cBbar}{{\bar{\cB}}}
\newcommand{\bBbar}{{\bar{\bB}}}
\newcommand{\bC}{{\bf C}}
\newcommand{\cbar}{{\bar{c}}}
\newcommand{\Cbar}{{\bar{C}}}
\newcommand{\Hbar}{{\bar{H}}}
\newcommand{\cL}{\mathcal{L}}
\newcommand{\bL}{{\bf L}}
\newcommand{\Lbar}{{\bar{L}}}
\newcommand{\cLbar}{{\bar{\cL}}}
\newcommand{\bLbar}{{\bar{\bL}}}
\newcommand{\cM}{{\cal M}}
\newcommand{\bM}{{\bf M}}
\newcommand{\Mbar}{{\bar{M}}}
\newcommand{\cMbar}{{\bar{\cM}}}
\newcommand{\bMbar}{{\bar{\bM}}}
\newcommand{\cP}{{\cal P}}
\newcommand{\cQ}{{\cal Q}}
\newcommand{\bU}{{\bf U}}
\newcommand{\bR}{{\bf R}}
\newcommand{\cW}{{\cal W}}
\newcommand{\bW}{{\bf W}}
\newcommand{\bZ}{{\bf Z}}
\newcommand{\Wbar}{{\bar{W}}}
\newcommand{\Xbar}{{\bar{X}}}
\newcommand{\cWbar}{{\bar{\cW}}}
\newcommand{\bWbar}{{\bar{\bW}}}
\newcommand{\abar}{{\bar{a}}}
\newcommand{\nbar}{{\bar{n}}}
\newcommand{\pbar}{{\bar{p}}}
\newcommand{\tbar}{{\bar{t}}}
\newcommand{\ubar}{{\bar{u}}}
\newcommand{\utilde}{\tilde{u}}
\newcommand{\vbar}{{\bar{v}}}
\newcommand{\wbar}{{\bar{w}}}
\newcommand{\phibar}{{\bar{\phi}}}
\newcommand{\Psibar}{{\bar{\Psi}}}
\newcommand{\bLambda}{{\bf \Lambda}}
\newcommand{\bDelta}{{\bf \Delta}}
\newcommand{\p}{\partial}
\newcommand{\om}{{\Omega \cal G}}
\newcommand{\ID}{{\mathbb{D}}}

\def\ep{\epsilon}
\def\de{\delta}

\def\wt{\widetilde}
\def\fracpd#1#2{\frac{\partial #1}{\partial #2}}
\def\pd#1#2{\frac{\partial #1}{\partial #2}}

\title{Reduction and integrability: a geometric perspective} 
\author{Jos\'e F. Cari\~nena\footnote{E-mail: jfc@unizar.es}\\
Departamento de F\'{\i}sica Te\'orica\\ 
Universidad de Zaragoza, 50009 Zaragoza, Spain\\
Real  Academia de Ciencias
Exactas, F\'{\i}sicas, Qu\'{\i}micas y Naturales de Zaragoza
}

\date{ }

\maketitle


\begin{abstract} 
A geometric approach to integrability and reduction of dynamical system is developed from a modern perspective. The main ingredients in such analysis are the infinitesimal symmetries and the tensor fields  that are invariant under the given dynamics.
Particular emphasis is given to the existence of invariant volume forms and the associated Jacobi multiplier theory,  and then the  Hojman symmetry theory is developed as a complement 
to Noether theorem and non-Noether constants of motion. The geometric approach to Hamilton-Jacobi equation is shown to be a particular example of the search for related field in a lower dimensional manifold.
\end{abstract}

\smallskip


\bigskip


\bigskip


\section{Introduction}

Systems of (may be partial) differential equations play a relevant r\^ole in the development of science and technology, and they quite often appear in many different branches of science ranging 
from pure mathematics to classical and quantum physics, control theory, economy, biology, etc.. For instance, the dynamical
evolution of deterministic systems  is described by systems of (may be time-dependent) first-order systems. However,
the solution of such systems is not an easy task.  The study of their integrability is a subject of significant
interest that has been an active field of research in recent years and appears very often in physics and mathematics
and with very different meanings. Here  by integrability of 
an autonomous system we mean that we are able to determine the general solution of the system. Such solutions can be searched in
 a specific family, for instance polynomial, rational functions or more generally, the so-called integrability by quadratures, that  means that it is possible to determine 
 the solutions by means of a finite number of algebraic operations, use of the Implicit Function Theorem and quadratures of some appropriate fundamental  functions, including rational, trigonometric and exponential functions, as well as their inverses.

To solve specific problems   reduction techniques are used. Students in physics relate the reduction process with infinitesimal symmetries
and  existence of constants of motion,  mainly in the framework of Lagrangian or
Hamiltonian mechanics. However
reduction processes can be developed in a more general framework, allowing therefore other interesting cases also relevant in physics,  and need not to
be related with symmetry properties but only aim to find simpler solvable systems
providing, at least partially, information on the solution of the original problem. In other words, given a difficult to solve problem  we can look
 for related simpler problems whose
solutions provide at least partial information on the original one. Sometimes we find geometric structures difficult to handle, for instance because
they have some kind of degeneracy. We look then for a related structure of the
same kind but easier to manage, for instance a non-degenerate one. 

The existence of additional compatible geometric structures, like symplectic or Poisson
structures may be useful in the search for solutions, and therefore
when the original problem admits some geometric structures, the reduction
procedure tries to preserve such structures. There is no systematic way of selecting the related reduced systems and sometimes
there are different possibilities. We generally assume that a system with less unknown variables or of
lower-order differential equations is simpler than other with more unknown variables
or of higher-order differential equations, but reduction processes from linear systems produce nonlinear interacting ones.
 
The simplest case is that of an autonomous system of first-order
differential equations (the higher-order case can be related to some first-order one)
\begin{equation}
\dot x^i=  X^i(x^1,\ldots,x^n)\ ,\qquad  i=1,\ldots,n, \label{autsyst}
\end{equation} 
while non-autonomous systems correspond to the case in which the functions $X^i$ also depend on the independent variable $t$. In order to incorporate holonomic constraints, 
such a system  (\ref{autsyst})  is geometrically interpreted in terms of a vector field $X$ in a 
$n$-dimensional manifold $M$ with 
a coordinate expression in a local chart   
\begin{equation}
X=\sum_{i=1}^nX^i(x^1,\ldots,x^n)\pd{}{x^i}\ ,\label{avf}
\end{equation} 
in such a way that its integral curves  are  the solutions of the given system. 
In this sense, integrability by quadratures of a vector field  means that you can determine its integral curves (i.e. the flow of $X$) by quadratures, i.e.
  by means of a finite number of algebraic operations and quadratures of some functions. In a general case
  the flows of the vector fields cannot be found   in an  explicit way  using fundamental  functions, and 
we are happy if, at least, we can express the solutions in terms of  quadratures. The purpose  of integrability is therefore the 
characterisation of systems admitting such type of solutions and it has being receiving a lot of attention. The answer is always based on the use of
 appropriated Lie algebras of vector fields containing the given vector field. 
 
 From the geometric viewpoint by reduction of the dynamics given by the vector field $X\in\mathfrak{X}(M)$
we mean to find a manifold $N$, a differentiable map $F:M\to N$ and 
a simpler  $F$-related vector field $Y\in\mathfrak{X}(N)$, i.e. such that $TF\circ X=Y\circ F$, and hence the integral curves of $Y$
are images under $F$  of the integral curves of  $X$. Sometimes it   remains the problem of the  reconstruction of the integral curves of  $X\in\mathfrak{X}(M)$
 from  the integral curves of   $Y\in\mathfrak{X}(N)$.

The first of the  two more important cases is  when $i:N\to M$ is an immersed submanifold of $M$ (later on we will study when $F:M\to N$ is a surjective  submersion).
Note that $X\in\mathfrak{X}(M)$  must be such that   $X_{|i(N)}$ is tangent to
$i(N)$ and therefore there exists a vector field $\bar X$ in the lower-dimensional manifold $N$ such that is
$i$-related to $X$ and whose  integral curves   provide us some integral curves of $X$. We can suppose that the $(n-r)$-submanifold $N$ is (at least locally) defined
by $r$ functions $F_1, \ldots,F_r$, which should be constants of motion for $X$
because of the assumed tangency condition.  Then a relevant  technique in the reduction process consists on determining $r$
functionally independent constants
of motion, because this  reduces the original problem to another problem of a $r$-parameter family   involving only
$(n-r)$ variables. This  provides us 
foliations such that the vector field $X$ is tangent to the leaves, and the problem is 
reduced  to those of a family of  lower dimensional ones,  one on each  leaf.

  More specifically,   if $r$ functionally independent 
constants of motion $F_1,\ldots,F_r$ (i.e. such that $dF_1\wedge\cdots\wedge dF_r\ne 0$) are known, it 
allows us to reduce the problem to that of a family of vector fields
 $\widetilde X_c$, with  $c\in\mathbb{R}^r$,   defined in the  $n-r$ dimensional 
submanifolds  $M_c$ given by  the 
level sets of the vector valued  function of rank $r$,
  $(F_1,\ldots,F_r):M\to \mathbb{R}^r$. 
  Of course the best situation is when $r=n-1$: the leaves 
are 1-dimensional, giving us the solutions to the problem, up to a reparametrisation.  

The second important case is when there exists a manifold $N$ and a surjective submersion  $F:M\to N$  such that
$X\in\mathfrak{X}(M)$ is a projectable vector field. The map $F$ defines an equivalence  relation in $M$, and then $N$ is the space of such equivalence
classes. Sometimes the starting point is the equivalence relation and   $F$
is the projection map. For instance, usually one considers a transformation  Lie  group $G$ of $M$ that is a
symmetry group for $X$ and the equivalence relation associated to the action:
 $N$ is the space of orbits, assumed to be a
differentiable manifold, the function $F$ being the natural projection of each
element of $M$ on its orbit. The symmetry condition implies that $X$ is a
projectable vector field.  
A particular example of symmetry  is given by the local flow of a 
vector field  $Y$ that is an infinitesimal  symmetry of $X$, i.e. a 
vector field  $Y\in\mathfrak{X}(M)$ such 
that  $[Y,X]=0$. Then  in a neighbourhood of a point where $Y$ is different from zero we can choose adapted coordinates,
$(y^1,\ldots,y^n)$, 
for which  $Y$ is written (Straightening out Theorem)
$Y=\partial/\partial {y^n}$. 
Then   $[Y,X]=0$ implies that $X$ has the form
$$
X=\bar X^1(y^1,\ldots,y^{ n-1})\,\pd{}{y^1}+\ldots +\bar X^{ n-1}(y^1,\ldots,y^{ n-1})\,\pd{}{y^{ n-1}}+
\bar X^ n(y^1,\ldots,y^{ n-1})\pd{}{y^ n}\ , 
$$
 and its integral curves are obtained by solving the system
$$\left\{\begin{array}{ccl}
{\displaystyle \frac{dy^i}{dt}}&=&\bar X^i(y^1,\ldots,y^{ n-1}), \qquad  i=1,\ldots , n-1,
\cr &&
\cr
{\displaystyle \frac{dy^ n}{dt}}&=& \bar X^ n(y^1,\ldots,y^{ n-1})\end{array}\ .
\right.
$$
We have  reduced the problem to a subsystem involving only the first 
 $ n-1$ equations, and once this has been solved,  the last equation is used to obtain  the function 
 $y^ n(t)$ by means of one quadrature. 
 Note however that the new coordinates $y^1,\ldots,y^{ n-1}$, are local constants of the motion for $X$ 
 and therefore we cannot find in an easy way such coordinates in a general case. Note also that the information provided by
 two different symmetry vector fields,  $Y_1$ and $Y_2$, cannot be used simultaneously in the general case,
because it is not possible to find local coordinates
 $(y^1,\ldots,y^n)$  such that
 $Y_1=\partial /\partial {y^{n-1}}$ and $Y_2=\partial /\partial {y^n}$,
unless that  $[Y_1,Y_2]=0$.

In other situations the equivalence relation is defined by a foliation
(involutive distribution) and the equivalence classes are the leaves.
The set of leaves is assumed to have a manifold structure.
If the original system has an invariance group we can consider invariant
foliations in such a way that leaves are preserved.

Other reduction  procedures for  systems can be developed when we  know  some particular solutions of 
them or related systems. For instance, one particular example is Riccati equation, of a fundamental importance both  in physics
(for instance factorisation of second order differential operators, Darboux transformations
and in general Supersymmetry in Quantum Mechanics) and in mathematics. 
It is well known   that if one particular solution of  Riccati equation is   known, we can find the general solution by means of  two quadratures, while
if two particular solutions are known, we can find  the general solution by means of just  one  quadrature, and finally 
if three  particular solutions are known we can explicitly write the solution  without any quadrature, by means of a  superposition function.
Actually there is a kind of systems admitting such a    superposition function, the so-called Lie systems, which 
appear very often in many problems in science and engineering (see \cite{CdL11,JdLCS} for a review). In the solution of such non-autonomous systems of first-order differential equations
we can use techniques imported from group theory, for instance Wei-Norman method \cite{WN1,WN2,CMN98},
and reduction techniques coming from the theory of connections.

Throughout the paper we shall take advantage of results in previous publications, where more details can be found. Section 2 is devoted to Lie  fundamental  result of integrability by quadratures of a system without recourse to any related invariant tensor field but only to properties of a Lie algebra of vector fields containing the given vector field. In section 3 the interest of invariant  tensorial fields 
is pointed out and the particular case of invariant (1,1)-tensor fields giving rise to recursion operators, also called symmetry generators,  and Lax equations is analysed. 
In Section 4, for  the sake of completeness we review some basic aspects of symplectic geometry,  and the classical  first Noether Theorems in both 
Hamiltonian and Lagrangian approach are summarised.
 The non-Noether constants of motion are derived in Section 5  from the existence of alternative invariant geometric structures for the  dynamical  vector field. As we want to add some comments on a different method of finding constants of motion, 
  as a prelude to deal in Section 7 with the theory of Hojman symmetry we develop  
  in Section 6  some remarkable points on   invariant volume forms and the 
 associated theory of Jacobi multipliers, a relevant  ingredient in integrability of classical mechanical systems.  Section 8 is a short summary of how to use the reduction theory   
to introduce Hamilton-Jacobi equation that appears here as a method of finding lower-dimensional systems related to a Hamiltonian one. 

\section{Lie integrability by quadratures}

We first summarise the pioneer work of Lie on integrability of some particular types of vector fields, without any recourse to the existence of additional 
compatible structures, but only using modern tools of algebra and geometry, in particular Lie algebras of symmetry vector
fields, and more specifically, solvable Lie algebras \cite{SLie91}.

\begin{lem}
If $n$ vector fields $X_1$,\ldots,$X_n$, which are linearly independent in each point of  an open set  $U\subset\mathbb{R}^n$, 
generate a solvable Lie algebra and  are  such that $[X_1,X_i]=\lambda_i\, X_1$ with $\lambda_i\in \mathbb{R} $,
then the differential equation  $\dot x=X_1(x)$ is solvable by quadratures in $U$.
\end{lem}
  
 {\sl Proof.-} We only prove here the simplest case $n = 2$. Then,  the derived algebra is 1-dimensional, and therefore the Lie algebra is solvable.
The differential equation can be integrated if we are able to find a first-integral $F$, i.e. $X_1F=0$,  such that $dF\ne 0$ in  an open set  $U$.  
In this case, for each real number $c\in\mathbb{R}$,  we can implicitly define one variable, for instance $x_2$,
 in  terms of the other one by $F(x_1,\phi(x_1))=k$,  and the differential equation determining the integral curves of $X_1$ is  in separate variables, i.e. integrable by quadratures.
 Note that the condition $[X_1,X_2]=\lambda_2\, X_1$, with $\lambda_2\in\mathbb{R}$, shows that if $F$ is a first-integral for  $X_1$ then $X_2F$ is a first-integral
  for $X_1$ too,  because $X_1(X_2F)=X_2(X_1F)+\lambda_2\, X_1F=0$. Note that  as $n=2$   there exists a 1-form $\alpha_0$, 
which is defined up to multiplication by a function, such that $i(X_1)\alpha_0=0$. Obviously as $X_2$ is linearly independent of $X_1$ at each point, $i(X_2)\alpha_0\ne 0$.  
We can see that the
1-form $\alpha=(i(X_2)\alpha_0)^{-1}\alpha_0$ satisfies the condition $i(X_2)\alpha=1 $, by definition of $\alpha$, 
and that  it is closed, because  $X_1$ and  $X_2$ generate $\mathfrak {X}(\mathbb{R}^2)$ and
$$d\alpha ( X_1,  X_2 ) =  X_1\alpha( X_2 )-X_2\alpha( X_1) + \alpha ( [ X_1,  X_2 ]) = \alpha ( [ X_1,  X_2 ]) = \lambda_2\, \alpha ( X_1) = 0.$$
Therefore, there exists, at least locally, a function $F$ such that $\alpha = dF$, and  the condition $i (X_1)\alpha = 0$  means that the function $F$ is
a first-integral for $X_1$.
\hfill$\Box$

A generalisation of these results was given in \cite{CFGR} where the vector fields $X_1,\ldots, X_n$ are assumed to close on a real Lie algebra,  
i.e. $[X_i,X_j]={\displaystyle \sum_{k=1}^n} c_{ij}\,^kX_k$, 
with $ c_{ij}\,^k\in \mathbb{R}$, 
and it was proved that if such Lie algebra is solvable and $A$ is an Abelian ideal, then the vector field $X$ is integrable for each vector field $X$ in the ideal 
$A$, and in particular, if the Lie algebra is nilpotent, any vector field of the Lie algebra is integrable. This result was extended
 in \cite{CFG} where it was proved that solvability of a Lie algebra of vector fields implies their integrability by quadratures.
 
 \section{Invariant tensor fields and integrability of a vector field}
 
 We have indicated how the knowledge of a first-integral, i.e. an invariant function, can be used to reduce the integrability of the vector field to a family of  lower
  dimensional problems. This is a particular example of a more general case, the r\^ole of tensor fields invariant  under a vector field $X$ on a $n$-dimensional manifold $M$ in its integrability 
  (see e.g. \cite{kozlov19}). For instance, invariant vector fields $Y$  give rise to infinitesimal symmetries of $X$, whose flows transform solutions into solutions. Moreover, if $F$ is a first-integral for $X$, then 
  $Y(F)$ is also a first-integral, because $\mathcal{L}_X(Y(F))=Y\left(\mathcal{L}_X(F)\right)=0$. The same property holds when   the vector field $Y$ is not a symmetry of the vector field $X$ but only of the 1-dimensional distribution spanned by $X$, i.e. there exists a
 function $h\in C^\infty(M) $ such that $[Y,X]=h\, X$, because then  $\mathcal{L}_X(Y(F))=Y\left(\mathcal{L}_X(F)\right)-h\, X(F)=0.$

   Later on, we will see  how to find new constants of motion
   when additional tensor fields enter 
  in the game. Of course some invariant multivector fields are also interesting but particularly the case of Poisson bivector fields is very relevant. Differential forms
   that are invariant under  
    $X$ give rise to absolutely integral invariants, a theory developed by Poincar\'e in  \cite{hpoinc}. In particular,  invariant 2-forms as for instance the relevant case in 
    mechanics of symplectic forms, are very relevant in the study of integrability of $X$ and Arnold-Liouville integrability is based on
     the existence of an appropriated invariant symplectic form. 
    Even if we have a $(1,1)$-tensor field $\mathcal{R}$ invariant under $X$, it may be used as a generator of symmetry, in the sense that if 
    $Y$ is a symmetry of $X$,  then $\mathcal{L}_X\mathcal{R}=0$  implies that $\mathcal{L}_X(\mathcal{R}(Y))=0$. Furthermore,  if we choose  in this case
     a local basis of vector fields   $\{ X_1,\ldots,X_n\}$, then  if the 
 (1,1)-tensor field $\mathcal{R}$ is 
invariant under  $X\in \mathfrak{X} (M)$ and the  matrices $A$ and  $B$ are the matrix 
representatives of  $\mathcal{R} $ and $\mathcal{L}_X$ in such a basis,  respectively, then 
the invariance condition $\mathcal{L}_X\mathcal{R}=0$ is written as
 \begin{equation}
\dot A= [B,A],  \label{laxeq}
\end{equation}
 where  $\dot A_i\,^j$ denotes  $X (A_i\,^j)$, because if   $ \mathcal{R} (X_i) = {\displaystyle \sum_{j=1}^n} A_i\,^jX_j $  and   
 $\mathcal{L}_X X_i={\displaystyle \sum_{j=1}^n}B_i\,^j X_j$, as we have 
 $       \mathcal{L}_X [ \mathcal{R} (X_i)] ={\displaystyle \sum_{j=1}^n}\mathcal{L}_X [A _i\,^jX_j ] = {\displaystyle \sum_{j=1}^n}\mathcal{L}_X(A_i\,^j)X_j +
  {\displaystyle \sum_{j,k=1}^n}A_i\,^jB_j\,^k X_k,$ and $ \mathcal{R} (\mathcal{L}_X X_i ) =  {\displaystyle \sum_{j,k=1}^n}B_i\,^jA_j\,^k  X_k,$
then,   taking into account that  $X$-invariance of   $\mathcal{R}$ is equivalent to   $  \mathcal{L}_X[ \mathcal{R} (X_i)] = \mathcal{R} [  \mathcal{L}_X(X_i) ]$,  we find that $ \mathcal{L}_X \mathcal{R}  =0$ is equivalent 
  to      $ \mathcal{L}_X(A_i\,^j)+{\displaystyle \sum_{k=1}^n}\ A_i\,^k B_k\,^j ={\displaystyle \sum_{k=1}^n} B_i \,^kA_k\,^j$, and therefore  to the matrix equation 
   (\ref{laxeq}), which is called Lax equation \cite{lax,Fi82,CI85a}. Note also that as the matrix representative of $ \mathcal{R}^2$ is $A^2$ and 
   $ \mathcal{L}_X\mathcal{R}  =0$ implies that $ \mathcal{L}_X\mathcal{R} ^k =0$, with $k\in \mathbb{N}$, we also have 
 \begin{equation}
\dot{A^k}= [B,A^k],\quad k=1,\ldots n.  \label{laxeqk}
\end{equation}
The importance of these equations is that, as for any pair of matrices the trace of the commutator is zero, we have that ${\displaystyle\frac d{dt}\textrm {Tr} A^k=0}$, and consequently a  $(1,1)$-tensor
 field $\mathcal{R}$ invariant under $X$ provides us $n$ constants of motion, in principle not all of them functionally independent, as indicated by   the Hamilton-Cayley 
 theorem. Another remark is that, as the coefficients of the characteristic equation $\det (A-\lambda I)=0$ can be reconstructed from the traces of powers 
 of $A$ (Le Verrier method of determining the characteristic
equation of a matrix, see e.g. \cite{wilk65}),  the roots of such  characteristic equation, the eigenvalues of $A$,  are constants of motion.

 The search of these invariant $(1,1)$-tensor fields $\mathcal{R}$ is not an easy task and use to come from the existence of alternative structures \cite{CI83,CR88,CFR13}, 
 the associated constants of motion being called non-Noether constants of motion (see.e.g. \cite{CI83,Cr83a,Ma83}), but they may have a 
 different origin \cite{Fi82,CI84a}. We will be back on this point in Section \ref{nonN}.
 
Another interesting case is when there exists a volume form  invariant under $X$. This case and an alternative method to find    first-integrals associated to 
symmetries of 1-dimensional distribution spanned by  the vector field $X$ will be analysed in Sections \ref{JMsection} and \ref{Hojmansym}.

\section{Invariant symplectic structures}\label{sympstr}

 We recall that $(M,\omega)$ is
 a symplectic manifold if  $M$ is a
 finite-dimensional differentiable manifold and $\omega$ is a nondegenerate 2--form
 which satisfies $d\omega = 0$ (that is, $\omega\in Z^2(M)$). The dimension of $M$ is
 necessarily even, $\dim M=2n$. For general results, reference textbooks are, for instance,
 \cite{AM} and~\cite{LM}.
 By nondegeneracy of $\omega$ we mean that 
  for every point $u\in M$ the map $\widehat\omega_u:T_uM \to T_u^*M$, given by:
 $   \langle \widehat\omega_u(v),v' \rangle= \omega_u(v,v')$ with $v,v'\in T_uM$,
 has a maximal rank, i.e. $(\omega)^{\wedge n}\ne 0$.
 Such a  map $\widehat \omega$ is a  base-preserving fibered 
 map, and hence   it induces a mapping between the linear spaces of  sections which, with a slight abuse of
 notation, we will also write $\widehat\omega:\mathfrak{X}(M)\to \bigwedge^1 (M)$.

        The following well-known result  completely characterises these symplectic 
        manifolds, from
the local point of view:

\begin{theorem}(Darboux)  Around each point $u \in M$ with $\dim M=2n$ there is a local
chart $(U,\phi)$ such that if $\phi = (q^1,\dots,q^n; p_1,\dots,p_n)$, then
$\omega|_U = {\displaystyle\sum_{i=1}^n} dq^i \wedge dp_i$.
\end{theorem}
\hfill$\Box$

Such coordinates are called Darboux coordinates.

Since closed, and in particular exact, 1-forms are distinguished elements in  $\bigwedge^1 (M)$, the corresponding vector fields are called locally Hamiltonian vector fields and Hamiltonian vector fields, respectively. If $H\in C^\infty(M)$, 
 the  Hamiltonian vector field  $X_H$ associated
 with the Hamiltonian $H$ is the unique vector field satisfying
 $ \widehat\omega(X_H) =i(X_H)\omega= dH$.
The set of Hamiltonian vector fields will be denoted $\mathfrak{X}_{{\rm H}}(M,\omega)$ and that of locally Hamiltonian vector fields,  $\mathfrak{X}_{{\rm LH}}(M,\omega)$, i.e. 
 $\mathfrak{X}_{{\rm LH}}(M,\omega) = \widehat\omega^{-1}(Z^1(M))$ and $ \mathfrak{X}_{{\rm H}}(M,\omega) = \widehat\omega^{-1}(B^1(M))$.
 Observe that  $\widehat\omega^{-1}$ is an isomorphism of real vector spaces.
   
The Cartan {\sl homotopy identity\/}, i.e. $\mathcal{L}_X\alpha = i(X)d\alpha +
 d(i(X)\alpha)$,  for any $\alpha\in  \bigwedge^p(M)$, shows that, given  $X\in\mathfrak{X} (M)$,  
 $\mathcal{L}_X\omega =0$ if and only if $i(X)\omega$ is a closed 1--form, i.e. 
 $X\in \mathfrak{X}_{{\rm LH}}(M,\omega)$.   In particular,    $\mathcal{L}_{X_H}\omega =0$.     In Darboux coordinates the
 Hamiltonian vector field $X_H$ corresponding to the function $H$ is given by
$$
        X_H = \sum_{i=1}^N\left(\pd H{p_i}\pd{}{q^i} - \pd H{q^i}\pd{}
{p_i}\right) \,,    
$$
and therefore, the equations determining its integral curves are similar to Hamilton 
equations.

 Remark that if $X,Y\in \mathfrak{X}_{{\rm LH}}(M,\omega) $ the  
 commutator $[X,Y]$ is a Hamiltonian vector field, with Hamiltonian
 $\omega (Y,X)$, because from  the relation 
 $i(X) \mathcal{L}_Y\alpha - \mathcal{L}_Y i(X)\alpha = i([X,Y])\alpha$,
 valid for any form $\alpha$, we obtain:
 $$ i([X,Y])\omega=  i(X)\mathcal{L}_Y\omega - \mathcal{L}_Y i(X)\omega = - \mathcal{L}_Y i(X)\omega =   
 - i(Y)d(i(X)\omega) - d(i(Y)i(X)\omega)\,,   
$$
and then,
\begin{equation}
i([X,Y])\omega = - d(\omega(X,Y))\,.\label{homom}
\end{equation}

 Consequently the set  $\mathfrak{X}_{{\rm LH}}(M,\omega)$ is a Lie algebra and  $\mathfrak{X}_{{\rm H}}(M,\omega)$ is an ideal in  $\mathfrak{X}_{{\rm LH}}(M,\omega)$.  
   
 As an important property, if $(M,\omega)$ is a symplectic manifold of dimension $2n$ we
 define the Poisson bracket of two functions $F,G  \in C^\infty(M)$ as being  the
 function $\{F,G\}$ given by:
 $$
 \{F,G\} = \omega(X_F,X_G) = dF(X_G) = - dG(X_F)\ .
 $$

In Darboux coordinates for $\omega$ the expression for $\{ F,G\}$ is 
 the usual one: 
 $$ \{ F,G\}  =\sum_{i=1}^N\left(\pd{F}{q^i}\pd G{p_i} -\pd{F}{p_i}
\pd G{q^i}\right) .$$

The aforementioned property (\ref{homom}) means that if  $F,G  \in C^\infty(M)$,  then we have that  $d\{F,G\} = -i ([X_F,X_G])\omega$, 
 i.e. $[X_F,X_G] =X_{\{ G,F\}} $.
 
 This  Poisson bracket  
$\{\cdot,\cdot\}$ is a skewsymmetric $\mathbb{R}$--bilinear map on $C^\infty(M)$ such that it
satisfies Jacobi's identity, as a consequence of $\omega $ being closed. In fact,
if $F,G,H\in C^\infty (M)$, 
$$
\aligned        (d\omega)(X_F,X_G,X_H)&= X_F\omega(X_G,X_H)- X_G\omega(X_F,X_H)+ X_H
\omega(X_F,X_G)\\
        &-\omega([X_F,X_G], X_H)+\omega([X_F,X_H], X_G)-\omega([X_G,X_H], X_F)\ ,
\endaligned
$$
and taking into account that  
$X_F\omega( X_G,X_H)= X_F\{G,H\}=\{\{G,H\},F\}$ 
and that  
 $\omega([X_F, X_G],X_H)$  can also be rewritten as 
$ \omega([X_F,X_G],X_H)=\omega(X_{\{G,F\}},X_H)=\{\{G,F\},H\}$,
as well as  the corresponding expressions for each cyclic 
reordering,  we find that
$$
     (d\omega) (X_F,X_G,X_H)=2[\{\{G,H\},F\}+ \{\{H,F\},G\}+ \{\{F,G\},H\}]\ .
$$
Finally, we recall that in each point of a  neighborhood of every point
 of $M$ 
the values of the set of 
Hamiltonian vector fields 
generate the tangent space, and therefore $d\omega=0$ if and only if 
Jacobi identity holds. Hence, the Poisson bracket endowes $C^\infty (M)$ with a real  Lie algebra
 structure, the Jacobi identity being a consequence of the closedness of $\omega$, and  the above mentioned property shows that $-\widehat \omega^{-1}\circ d: C^\infty (M\,\{\cdot,\cdot\}\to  \mathfrak{X}_{{\rm H}}(M,\omega) $ 
 is a Lie algebra homomorphism.
 
Given a Hamiltonian system $(M,\omega,H)$ one usually look for vector fields whose flows are symplectomorphisms and such  that are 
  also symmetries of $H$ and, therefore, symmetries of $X_H$. Then for each $F\in C^\infty(M)$, the relation 
  $X_HF=\{F,H\}=-X_FH$
 shows that $X_F$ is a symmetry of $H$ if and only if $F$ is a constant of motion (a result usually known as Noether's theorem). 
It is to be remarked that in the case of a Hamiltonian dynamical system the constants of motion $F$  are 
playing a double r\^ole in the reduction process, either as constants of motion, or as generating 
infinitesimal symmetries $X_F$, because if 
$F$ is a constant of the motion, $\{F,H\}=0$,
then the  Hamiltonian vector field  $X_F$ is a symmetry of  $H$.

Moreover, both $X_H$ and  $X_F$ are tangent to the level sets of
$F$, because $X_H F=\{F,H\}=0$, and  $X_FF=\{F,F\}=0$. 
The restriction 
of $X_F$ on each leaf of the foliation $\mathcal{F}_H$ whose leaves are the level sets of $F$
can be used to determine adapted coordinates in such a way that 
the problem of determining the integral curves of $X_H$ 
is reduced not just in one  but in two degrees of freedom.

In order to be able to use simultaneously the 
information given by  different constants of motion, $F_1,\ldots,F_r$, it is 
sufficient that $\{F_i,F_j\}=0,\,\forall i,j=1,\ldots, r$, because
$[X_{F_i},X_{F_j}]=X_{\{F_j,F_i\}}$. If the condition is satisfied, then  
$[X_{F_i},X_{F_j}]=0$, for any pair of indices $i$ and $j$, and we can find adapted coordinates such that $X_{F_i}=\partial /\partial y_ i$, for $i=1,\ldots, r$. 
The condition is not necessary because 
if, for instance,  $\{F_i,F_j\}$ is constant for each pair of indices, then it is also true that $[X_{F_i},X_{F_j}]=0$.

The simplest case will be that of Hamiltonian systems in a space of dimension
 $N=2\,n$ for which such $n$ constants of motion, $F_1,\ldots,F_n$, in involution
  are known: they are called completely integrable systems. 

We are now ready to  recall the notion of Liouville-Arnold integrability: 

\begin{defn}  
The Hamiltonian dynamical system $(M,\omega,H)$, with $\,\dim M=2n$,  is said to be completely integrable if there exists 
a set of $n$
functions $\{ F_j\mid j=1,\ldots,n\}$, where $F_j\in C^\infty(M)$, with $F_1=H$, which are
constants of the motion, i.e.
$$
\frac{dF_k}{dt}=\mathcal{L}_{X_H} F_k=\{ F_k,H\}=0, \qquad \forall k=2,\ldots,n, 
$$
that are functionally independent, i.e.  $dF_1\land \cdots \land dF_n\neq 0$,
and  are pairwise  in involution, i.e.
$$
\{ F_k, F_j\}=0 ,\qquad \forall j,k=1,\ldots n.
$$
\end{defn}
A completely integrable system that admits more than $n$ functionally independent constants of motion is said to be superintegrable and when the number is the maximum number, $2n-1$, 
the system is called maximally superintegrable \cite{evans}. Many of such systems can be found in the physics literature (see e.g. the recent papers
  \cite{CRS21a,CRS21b} and references therein).

The two main examples of Hamiltonian dynamical systems are those of Hamiltonian systems on the cotangent bundle $T^*Q$ of a manifold
 $Q$ and those defined by  regular Lagrangians on the tangent bundle $TQ$. In fact the cotangent bundle $\pi:T^*Q\to Q$ is endowed with a canonical
  1-form $\theta\in \bigwedge^1(T^*Q)$ defined by $\theta_\alpha= \pi ^*_\alpha\alpha$, for all covectors  $\alpha$ in $Q$. Then $\omega=-d\theta$ is a symplectic form on $T^*Q$. It is to be remarked that if $\alpha$ is a 1-form in $Q$, then as $\alpha^*\theta=\alpha$ we have that    $\alpha^* \omega=\alpha^*(-d \theta)=-d(\alpha^* \theta)=-d\alpha$. 
  
  The geometric framework for the study of Lagrangian mechanics is that of tangent bundles \cite{MC81,MC83,CIMM}. The tangent bundle $\tau:TQ\to Q$ is characterised by two geometric tensors, the vertical endomorphism  
$S$, a (1,1)-tensor field on $TQ$, also called  tangent structure, which satisfies $\text{Im}\, S=\ker S$ and an integrability condition, the vanishing of the Nijenhuis tensor, $N_S=0$,  and the Liouville vector field $\Delta$ generating dilations along fibres in $TQ$ \cite{CT}. 

  If $(U,\varphi)$ is  a local chart  on $Q$ and ${\rm pr}^i:\mathbb{R}^n\to\mathbb{R}$ are the natural projections on the $i$-th-factor and $q^i={\rm pr}^i\circ \varphi$ we define the coordinate system   $(U,q^1,\ldots,q^n)$ on $Q$, and the corresponding chart in \,  $\mathcal{U}=\tau^{-1}(U)$ given by $(\mathcal{U},\varphi,\varphi_*)$,  defines a  coordinate system  $(q^1,\ldots,q^n,v^1,\ldots,v^n)$ on the open set  $\mathcal{U}=\tau^{-1}(U)$  of $TQ$.
 Correspondingly, we  consider the coordinate basis of $\mathfrak{X}(U)$ usually denoted $\{\partial/\partial q^j\mid j=1,\ldots ,n\}$ and its dual basis for $\bigwedge^1(U)$, $\{dq^j\mid j=1,\ldots ,n\}$. 
   Then a  vector $v$ in a point $q\in U$  is $v={\displaystyle \sum_{i=1}^n}v^j\,(\partial/\partial q^j)_{|q}$, i.e. $v^i=dq^i(v)$, while a  covector $\alpha $ in a point $q\in U$  is $\alpha={\displaystyle \sum_{j=1}^n}p_j\, dq^j\,_{|q}$, with $p_i=\alpha((\partial/\partial q^i)_{|q})$.

 With this notation the  coordinate expressions of the  vertical endomorphism $S$ and the Liouville vector field $\Delta$,
 are  \cite{MC81,MC83}:
\begin{equation}
 S(q,v)=\sum_{i=1}^n\pd{}{v^i}\otimes dq^i, \qquad 
   \Delta(q,v)=\sum_{i=1}^nv^i\pd{}{v^i}.\label{VESLvf}
   \end{equation}
    Similarly we can introduce a coordinate system $(q^1,\ldots,q^n,p_1,\ldots,p_n)$ in the open set $\bar{\mathcal{U}}=\pi^{-1}(U)$ of the cotangent bundle $\pi:T^*Q\to Q$
   and   the coordinate expression of the 1-form  $\theta$ is  
   \begin{equation}
\theta(q,p)=\sum_{i=1}^np_i\,  dq^i, \label{ectheta}
   \end{equation}
with shows that $\omega=-d\theta$ is a canonical symplectic structure on $T^*Q$.

The vector field $\Delta$ and the tensor field $S$ can be used to select a special kind of vector fields whose integral curves are given by lifting solutions of a second order differential equations. These vector fields called  second order differential equation vector fields  $D\in\mathfrak{X}(TQ)$ (hereafter shortened as SODE vector fields) are characterised by $S(D)=\Delta$.

Recall  also that given a  Lagrangian $L\in C^\infty(TQ)$ we can define a 1-form $\theta_L=dL\circ S$  and the exact  2-form $\omega_L=-d\theta_L$. When $\omega_L$ 
is regular the Lagrangian $L$ is said to be regular and then  the dynamics is given by the uniquely defined  SODE 
vector field $\Gamma_L$ such that 
\begin{equation}
i(\Gamma_L)\omega_L=dE_L\Longleftrightarrow \mathcal{L}_{\Gamma_L} \theta_L-dL=0, \label{dyneq}
\end{equation}
where the energy function $E_L\in C^\infty(TQ)$ is defined by $E_L=\Delta (L)-L$. Moreover, this implies that  $\mathcal{L}_{\Gamma_L} \omega_L=d(i(\Gamma_L)\omega_L)=0$

In usual local tangent bundle coordinates the expressions of $\theta_L$ and $E_L$ are
\begin{equation}
\theta_L=\sum_{i=1}^n\pd L{v^i}\, dq^i,\qquad E_L= \sum_{i=1}^nv^i\pd L{v^i}-L,\label{lcetetaE}
\end{equation}
while that of $\omega_L$ is:
\begin{equation}
\omega_L=\sum_{i,j=1}^n\pd{^2 L}{{ q}^j\partial  v^i  } d { q}^i\wedge dq ^j+\sum_{j,k=1}^n\pd{^2 L}{ v^k \partial  v^j} 
d q^j\wedge dv ^k. \label{lceomegaL}
\end{equation}

It may be of  be interest under what conditions two regular Lagrange functions $L,L'\in C^\infty(TQ)$ in $TQ$ produce the same symplectic structure and the same energy function, i.e. $L_0=L-L'$ 
is such that $\omega_{L_0} \equiv0$ and $E_{L_0}=0$. 

Now, if $\alpha$  is a 1-form on $Q$, $\alpha\in\bigwedge^1(Q)$,  then $\widehat\alpha $ denotes  the function $\widehat\alpha \in C^\infty (TQ)$ 
defined by $\widehat\alpha(u) = \alpha_{\pi(u)}(u)$. If the local coordinate expression of a  1-form $\alpha$ is $\alpha={\displaystyle\sum_{i=1}^n}\alpha_i(q)\, dq^i$, then 
$\widehat\alpha(q,v)={\displaystyle\sum_{i=1}^n}\alpha_i(q)\, v^i$. Consequently $\Delta \widehat\alpha=\widehat\alpha$, because, by definition $\widehat\alpha$ is a homogeneous of degree one
function. On the other hand, as 
$$(d\widehat\alpha\circ S)\left(\pd{}{q^i}\right) =\alpha_i(q),\qquad  (d\widehat\alpha\circ S)\left(\pd{}{v^i}\right) =0,
$$
we see that $d\widehat\alpha\circ S=\tau^*\alpha$.

Then, 
a function $ L_0\in C^\infty(TQ)$ is such that $\omega_{L_0} \equiv0$ if and only if there exist a
closed 1-form $\alpha\in Z^1( Q)$  and a function $h \in C^\infty(Q)$  such that $L_0 = \widehat \alpha + h\circ \tau$ (see e.g. \cite{CI85b}). But as for such a function $L_0$ we have 
that $E_{L_0}=-h$, we see that  $L,L'\in C^\infty(TQ)$ in $TQ$ produce the same Hamiltonian dynamical system, i.e. $\omega_L=\omega_{L'}$ and $E_L=E_{L'}$, if and only if there exists a closed 1-form $\alpha\in Z^1(Q)$ such that $L'=L+
\widehat\alpha $, and both Lagrangians are then said to be gauge equivalent. 

It is also possible to identify a complete lift vector field $X^c\in \mathfrak{X}(TQ)$ that is a symmetry of the Hamiltonian dynamical system $(TQ,\omega_L,E_L)$ in terms of 
symmetries of $L$ itself. Recall that if $X\in   \mathfrak{X}(Q)$, its complete lift, denoted $X^c$, is the vector field on $TQ$ whose flow is given by $\phi_{t*}$ where 
$\phi_{t}$ is the local flow of the vector field $X$. If the local coordinate expression of the vector field $X$ in a chart of $Q$  is $X(q)= {\displaystyle\sum_{i=1}^n}X^i(q) \partial/\partial q^i$, then the corresponding
expression for $X^c$ in  the associated chart of $TQ$ is $X^c(q,v)= {\displaystyle\sum_{i=1}^n}X^i(q) \partial/\partial q^i+{\displaystyle\sum_{i,j=1}^n}(\partial X^i/\partial q^j)v^j \partial/\partial v^i$.

It is also to be remarked that if  $\Phi\in{\rm Diff}(TQ) $ is a lift of a diffeomorphism $\varphi$ of the manifold $Q$, then $\Phi^*\theta_L=\theta_{\Phi^*L}$ as well as 
 $\Phi^*E_L=E_{\Phi^*L}$ as a consequence of $\Phi_*\Delta=\Delta$ and $[\Phi_*,S]=0$, because 
$$\Phi^*\theta_L=\Phi^*(dL\circ S)=dL\circ S\circ \Phi_*=dL\circ \Phi_*\circ S=d(L\circ \Phi)\circ S=\theta_{\Phi^*L},
$$
while, as $\Delta$ is $\Phi$-related with itself, $(\Delta L)\circ \Phi=\Delta(L\circ \Phi)$, we have 
$$\Phi^*E_L=\Phi^*(\Delta L)-\Phi^*L=\Delta(\Phi^*L)-\Phi^*L=E_{\Phi^*L}.
$$
At the infinitesimal level this means that if $X^c\in \mathfrak{X}(TQ)$ is the complete lift of $X\in \mathfrak{X}(Q)$ we have that
$$\mathcal{L}_{X^c}\theta_L=\theta_{X^cL},\qquad X^cE_L=E_{X^cL}.
$$
All these properties can be used to derive the Noether's theorem in the Lagrangian formalism:
\begin{theorem} (Noether) If the vector field $X\in \mathfrak{X}(Q)$ is such that there exists a function $h\in C^\infty(Q)$ such that $X^cL=\widehat{dh}$, then the function 
$f=i(X^c)\theta_L-\tau ^*h$ is a constant of motion.
\end{theorem} 
{\sl Proof.- } First, the vector field $X^c$ is Hamiltonian because 
$$\mathcal{L}_{X^c}\theta_L=\theta_{X^cL}= \theta_{\widehat{dh}}= \tau^* (dh)=d(\tau^*h),
$$ and then 
$$
i(X^c)d\theta_L+d(i(X^c)\theta_L)=d(\tau^*h)\Longleftrightarrow i(X^c)\omega_L=d((i(X^c)\theta_L-h)=df.
$$
Moreover, $X^cE_L=E_{X^cL}= E_{\widehat{dh}}=0$, and then if $\Gamma_L$ is the dynamical vector field determined by  $L$, i.e. $i(\Gamma_L)\omega_L=dE_L$,  we have
$$\Gamma_L f=\{f,E_L\}= -\{E_L,f\}=-X^cE_L=0.$$
\hfill$\Box$

Particularly interesting cases are the geodesic Lagrangians  defined through a Riemann structure $g$ by $L=\frac 12 g(v,v)$ (see e.g. \cite{TA10})   and the  so called natural Lagrangians defined by means of  a Riemann structure $g$ and a function $V$ in $Q$ as follows: $L=\frac 12 g(v,v)-V(q)$.

\section{Non-Noether constants of motion} \label{nonN}

Given a locally Hamiltonian vector field $X$  in a $2n$-dimensional symplectic manifold $(M,\omega)$, if $\omega'$ is another closed 2-form invariant under $X$, then we can consider the   pencil of invariant under $X$  closed 2-forms $\{ \omega ' - \lambda \,\omega \mid \lambda\in\mathbb{R} \}$, which defines a function $f \in\mathbb{R}\times  M \to\mathbb{R}$
such that    $(\omega ' - \lambda\, \omega)^{\wedge n} = f \,(\omega)^{\wedge n}$,  called characteristic function of the pencil,  and that,  by construction,  is a constant of the motion for each value of $\lambda$, because from  $\mathcal{L}_X ( \omega' - \lambda\, \omega ) ^{\wedge n} = 0$ we find  $\mathcal{L}_Xf=0$.   Moreover, the function $f$ is a polynomial function in $\lambda$ whose coefficients are constants of the motion. 
Note also that $\widehat \omega^{-1}\circ \widehat \omega':\mathfrak{X}(M)\to \mathfrak{X}(M)$ is $C^\infty(M)$-linear and define then a $(1,1)$-tensor field that is $X$-invariant 
  and such  that  the  characteristic polynomial  of such  composition coincides with the  characteristic function of the pencil, and therefore the mentioned constants of the motion for 
  $X$ coincide with those associated to the invariant $(1,1)$-tensor field  $\mathcal{R}=\widehat \omega^{-1}\circ \widehat \omega'$, and hence they are related to the traces of different powers of $\mathcal{R}$.

The point is how to find such a $X$-invariant 2-form $ \omega '$. We mention now two particular cases. First, if a vector field $Y$ is such that $[Y,X]=0$ but it is not  
a locally Hamiltonian vector field, then  $\omega'= \mathcal{L}_Y ( \omega)$ is a $X$-invariant closed 2-form, because 
 $ \mathcal{L}_X( \mathcal{L}_Y ( \omega))= \mathcal{L}_Y( \mathcal{L}_X ( \omega))=0$.
On the other side,  if   $\phi\in{\rm Diff}(M) $ is a noncanonical symmetry of the Hamiltonian vector field defined by the Hamiltonian dynamical system $(M,\omega,H)$,  from the  relation $i(\phi_*(X_H))((\phi^{-1})^*(\omega))=d((\phi^{-1})^*H)$, we see that if $\phi_*(X_H)=X_H$ then $i(X_H)((\phi^{-1})^*(\omega))=d((\phi^{-1})^*H)$, and therefore $\mathcal{L}_{X_H}((\phi^{-1})^*(\omega))=0$, and equivalently $\mathcal{L}_{X_H}(\phi^*(\omega))=0$. As it was assumed that 
 $\omega'=\phi^*(\omega)\ne \omega$, we can choose such a $X$-invariant closed 2-form $\omega'$ to define, together with $\omega$,  the pencil. We also recall that  
$\phi\in{\rm Diff}(M) $ is said to be a canonoid transformation of the Hamiltonian dynamical system $(M,\omega,H)$ when $\phi_*(X_H)$ is also
a Hamiltonian vector field, or equivalently when $X_H$  is Hamiltonian with respect to the
transformed 2-form $\phi^*(\omega)$, and in this case  $ \mathcal{L}_{X_H}(\phi^*(\omega))=0$
\cite{CR88,CFR13}, i.e. we can also choose  $\omega'=\phi^*(\omega)$ as invariant $X_H$-form.

  \section{Invariant volume forms and Jacobi multipliers}\label{JMsection}
  A particularly interesting case with many applications not only in the theory of differential equations but in classical mechanics is that of volume forms invariant
   under a given vector field $X$ on an oriented manifold $(M,\Omega)$. Each volume form is of the form $R\, \Omega$ with $R\in C^\infty(M)$ and the invariance condition is 
   $\mathcal{L}_X(R\, \Omega)=0$,  and if we take into account that $\mathcal{L}_X(R\, \Omega)=d(i(X)(R\, \Omega))=d(i(R\, X)(\Omega))=\mathcal{L}_{R\,X}(\Omega)$ we see that 
   the invariance condition of $R\, \Omega$ under $X$  is equivalent to   the invariance condition  of $ \Omega$ under $R\, X$. Remark also that for each vector field $X\in\mathfrak{X}(M)$ the volume form $\mathcal{L}_X \Omega$ is proportional to $\Omega$,  and the proportionality function   is called divergence of $X$, i.e.
   \begin{equation}
\mathcal{L}_X(\Omega)={\rm div}(X)\,\Omega\,,  \label{defdiv}
\end{equation}  
and that if local coordinates are chosen such that $\Omega=dx^1\wedge \cdots\wedge dx^n$ and  the local expression of $X$ is $X={\displaystyle \sum_{i=1}^nX^i\pd{}{x^i}}$, then the local expression of ${\rm div} (X)$ is: 
 $${\rm div} (X)=\sum_{i=1}^n  \pd{X^i}{x^i}.
 $$
 Vector fields $X$ such that ${\rm div}(X)=0$ are called divergence-free vector fields and enjoy interesting properties.
 
 Recalling the properties of Lie derivatives,   we find that, as for each $f\in C^\infty(M)$, $\mathcal{L}_{f\,X}( \Omega)=\mathcal{L}_{X}(f\, \Omega)=f\, \mathcal{L}_X (\Omega)+(\mathcal{L}_Xf) \,\Omega$, using the definition (\ref{defdiv}) of ${\rm div} (X)$, we have that
\begin{equation}
{\rm div}(fX)=f\,{\rm div}(X)+X(f)\,. \label{divprod}
\end{equation}
 The nonvanishing  functions $R$ such that $\mathcal{L}_X(R\, \Omega)=\mathcal{L}_{R\, X}(\Omega)=0$, i.e. $ {\rm div}(R\, X)=0$,  are called Jacobi multipliers, and (\ref{divprod}) 
shows that the nonvanishing  function $R$ is a  Jacobi multiplier if and only if (see \cite{CS21,CFN21} and references therein)
\begin{equation}
R\,{\rm div}(X)+X(R)=0 \Longleftrightarrow {\rm div}(X)+X(\log R)=0\,. \label{JMcond}
\end{equation}

Locally defined Jacobi multipliers, obtained as particular solutions of (\ref{JMcond}), always exist, but in some particular cases global solutions, giving rise to   invariant volume forms, do not exist \cite{FGGPLA225,FGNM15}.

It is also to be remarked that it follows from the relation  $\mathcal{L}_X(R\, \Omega)=d(i(X)(R\, \Omega))=d(i(R\, X)\Omega)$  that the function  $R$ is a Jacobi multiplier for $X$ in the oriented manifold $(M,\Omega)$ iff and only if $f\,R$   is a Jacobi multiplier for $X$  in the oriented manifold 
$(M,f^{-1}\,\Omega)$, for each positive function $f$.
 
 As an interesting case we can consider a SODE vector field $\Gamma\in\mathfrak{X}(TQ)$ which admits a Lagrangian formulation $L$, 
 i.e. there exists a regular Lagrange  function  $L\in C^\infty(TQ)$,
 such that $i(\Gamma)\omega_L=dE_L$, or equivalently  $\mathcal{L}_{\Gamma}\theta_L=dL$. Remark first that $TQ$ is orientable and a local chart of $TQ$ induced from one in $Q$
induces an orientation. As $\mathcal{L}_{\Gamma}\omega_L=0$, we see that the volume form $ (\omega_L)^{\wedge n}$ is $\Gamma$-invariant and therefore if a volume form was previously 
  fixed, there will be a Jacobi multiplier $R$ such that $(\omega_L)^{\wedge n}=R\,\Omega$. If, for instance, $\Omega$ is the volume form determined by a tangent bundle local chart, $\Omega=dq^1\wedge\cdots\wedge dq^n\wedge dv^1\wedge\cdots\wedge dv^n$,  we obtain, according to (\ref{lceomegaL}),  that the determinant of the Hessian matrix $W$  with elements $W_{ij}=\partial ^2L/\partial v^i\partial v^j$ is a Jacobi multiplier, because $(\omega_L)^{\wedge n}$ is a real multiple of $\det W\, dq^1\wedge\cdots\wedge dq^n\wedge dv^1\wedge\cdots\wedge dv^n$.

\section{Hojman symmetry and  constants of motion}\label{Hojmansym}

In the preceding sections \ref{sympstr} and \ref{nonN}  we have first summarised the usual way of searching first-integrals by means of infinitesimal  symmetries via the  first Noether theorem and 
presented then a second procedure for  finding non-Noether constants of motion,  which is not based on symmetries but on the existence of alternative geometric structures for the  description of the vector field, what leads to the existence of a recursion operator.  We mention next a third approach  started by Hojman \cite{SH92a} and Gonz\'alez-Gasc\'on in \cite{FGGJPA27} and that it  is becoming more and more important during   the last years for its applications in $f(R)$-gravity and FRW cosmology \cite{DGR20,DGR21,WZLZ15,WZLZ16,CR13,PC15,PLC16}. 

The main result is very general and a direct consequence of a simple geometric relation that when considering particular cases contains 
many results scattered in the physics literature. So, different relations among vector fields and their divergences can be used to establish first-integrals and integral invariants for vector fields in a manifold $M$. 

The following geometric relation plays a fundamental r\^ole:  If $X, Y$ is an arbitrary pair of vector fields in an oriented manifold $(M,\Omega)$, then
\begin{equation}
\mathcal{L}_X({\rm div\,} (Y))-
\mathcal{L}_Y({\rm div\,} (X))={\rm div\,} ([X,Y]), \label{reldosdiv}
\end{equation}
because as  the Lie derivatives of a $p$-form $\alpha$  satisfy the relation 
 \begin{equation}
\mathcal{L}_X(\mathcal{L}_Y\alpha)-\mathcal{L}_Y(\mathcal{L}_X\alpha)=\mathcal{L}_{[X,Y]}\alpha, \label{LXdg0}
\end{equation}
and in the particular case  $\alpha=\Omega$, we have
\begin{equation}
\mathcal{L}_X(\mathcal{L}_Y\Omega)-\mathcal{L}_Y(\mathcal{L}_X\Omega)=\mathcal{L}_{[X,Y]}\Omega, \label{LXdgO}
\end{equation}
and from difference of $\mathcal{L}_X(\mathcal{L}_Y\Omega)=\mathcal{L}_X({\rm div}(Y)\,\Omega)=\mathcal{L}_X({\rm div}(Y))\,\Omega+{\rm div}(Y)\,{\rm div}(X)\,\Omega$ and the corresponding 
relation  $\mathcal{L}_Y(\mathcal{L}_X\Omega)=\mathcal{L}_Y({\rm div}(X)\,\Omega)=\mathcal{L}_Y({\rm div}(X))\,\Omega+{\rm div}(X)\,{\rm div}(Y)\,\Omega$ we obtain (\ref{reldosdiv}).

As a consequence of this relation (\ref{reldosdiv}) we see that 
 if the  vector field $X\in\mathfrak{X}(M)$ in an oriented manifold $(M,\Omega)$  is divergence-free, 
then if the vector field $Y$ is an infinitesimal symmetry of $X$, i.e. $[X,Y]=0$, we have that $\textrm{div\,}(Y)$
is a constant of the motion for $X$.
Something similar happens when  the vector field $Y$ is an infinitesimal symmetry of 
the 1-dimensional distribution generated by $X$, i.e. there exists a function $h$ such that $[Y,X]=h\, X$, because then  in this case 
   the function $\textrm{div\,}(Y)+h $
is a constant of the motion for $X$. Actually, 
if ${\rm div\,}(X)=0$, then $\mathcal{L}_X\Omega=0$, and hence,  as $[X,Y]=-h\, X$, 
$$\mathcal{L}_X(\mathcal{L}_Y\Omega) =\mathcal{L}_Y(\mathcal{L}_X\Omega)+\mathcal{L}_{-h\,X}\Omega=-\mathcal{L}_X(h\,\Omega)=-X(h)\,\Omega,
$$
and then,  from 
$$\mathcal{L}_X(\mathcal{L}_Y\Omega) =\mathcal{L}_X({\rm div\,}(Y)\,\Omega)=\mathcal{L}_X({\rm div\,}(Y))\Omega,
$$
we obtain that $\mathcal{L}_X({\rm div\,}(Y)+h)=0$, and therefore the following function $I$ is a constant of the motion for $X$.
\begin{equation}
I={\rm div\,}(Y)+h.\label{fiIh}
\end{equation}

If the vector field is not divergence-free, remark that for any nonvanishing function $R$ the constants of motion of $X$ coincide with those of $R\, X$, and in particular if $R$ is a 
Jacobi multiplier for $X$ the vector field $\bar X=R\, X$ is such that when $[Y,X]=h\, X$, we have that  $[Y,\bar X]=\bar h\, \bar X$ with $\bar h=(Y(R)/R)+h$, and hence we have the constant of motion for $\bar X$, and therefore for $X$,  given by 
\begin{equation}
I={\rm div\,}(Y)+\bar h={\rm div\,}(Y)+Y(\log R)+h.\label{fiIh2}
\end{equation}

These general results can be applied to specific examples, and we recover as particular examples many   previously found constants of motion. For instance one can apply the 
general theory to  both Hamltonian and Lagrangian formulations of autonomous systems, or generic second order differential equations. In the particular case of a 
system on an Euclidean configuration space admitting a Lagrangian formulation we mentioned at the end of the preceding section that a Jacobi multiplier is given by the determinant of the Hessian matrix $W$, 
and this explains the result obtained by Lutzky \cite{Lutzky95} as a particular case of   the constant of motion given by (\ref{fiIh2}).  In the more general case of $Q$ being a $n$-dimensional manifold, a local chart of coordinates $(x^1,\ldots,x^n)$  for $Q$  induces a local chart of coordinates in its tangent bundle, denoted  $(x^1,\ldots,x^n, v^1,\ldots,v^n)$, as indicated in Section \ref{sympstr}, and an 
associated  volume element  in such a chart given by 
$\Omega=dx^1\wedge\cdots\wedge dx^{n}\wedge dv^1 \wedge\cdots\wedge dv^{n}$. If the dynamical vector field $\Gamma$ is given in such coordinates by  
\begin{equation}  \Gamma=\sum_{i=1}^n \left(v^i \pd{}{x^i}+F^i(x,v) \pd{}{v^i}\right)
\label{SODEvf}\end{equation}
 is determined by the Lagrangian $L$, i.e. $i(\Gamma)\omega_L=dE_L$, which implies  $\mathcal{L}_\Gamma\omega_L=0 $ (see e.g \cite{CIMM}),  it has been shown
 (see e.g. \cite{CS21} and references therein) that 
 a particular Jacobi multiplier for $\Gamma$ with respect to the volume form $\Omega$ is  given by the determinant of the Hessian matrix $W$ in the velocities, with elements 
$W_{ij}=\partial^2 L/{\partial v^i\partial v^j}$.  Actually, from $\mathcal{L}_\Gamma\omega_L=0 $, we see that  $(\omega_L)^{\wedge n}$ is an invariant volume under $\Gamma$, and the proportionality factor of $(\omega_L)^{\wedge n}$ and $\Omega$ is a real multiple of $\det W$.  Therefore, the constant of motion  obtained in \cite{Lutzky95} is just  a particular case of the expression  (\ref{fiIh2}) for $R$ equal to the 
determinant of the Hessian matrix $W$. 
As far as 
nonautonomous systems are concerned,  the dynamical vector fields must be  replaced by 1-dimensional distributions and therefore the more general condition $[Y,X]=h\, X$, 
which means that the vector field $Y$ preserves the 1-dimensional distribution generated by $X$, is relevant in the context of symmetry for such systems. 
As in the autonomous cases,  we can study particular examples of non-autonomous systems of  first-order differential equations \cite{SH92a, FGGJPA27}  but also of Hamiltonian 
systems as in \cite{GW14},  or even  non-autonomous systems of second-order differential equations  \cite{SH92a, FGGJPA27,FGGLNC19,ZCh}, and in particular 
 systems admitting a Lagrangian  formulation \cite{Lutzky95,ZCh}. For a recent geometric presentation of all these results see e.g. \cite{CR21}.

\section{Hamilton-Jacobi equation as a reduction procedure}

As a last example of the reduction theory we analyse  the well-know Hamilton-Jacobi equation from a geometric perspective. There is a very recent  nice review paper on the subject  \cite{NRRM9} based on 
\cite{CGMRM06} to which I refer for more general results and   only  the reduction technique  for the particular case of a non-autonomous Hamiltonian systems  in the phase space $T^*Q$ is presented here. 
Recall  that 1-forms on $Q$ are sections of the cotangent bundle $\pi_Q:T^*Q\to Q$ while vector fields in $Q$ are sections for the projection map $\tau_Q:TQ\to Q$ of the tangent bundle. 
Then, given a Hamiltonian dynamical system $(T^*Q,\omega,H)$ which defines a Hamiltonian vector field $X_H\in\mathfrak{X}(T^*Q)$ by the relation $i(X_H)\omega=dH$, where $\omega$ is the canonical symplectic structure on  $T^*Q$,  the aim is to find a vector field
$Z\in\mathfrak{X}(Q)$ and a    1-form $\alpha\in \bigwedge^1(T^*Q)  $ such that $X_H\circ \alpha= T\alpha\circ Z$. Such a pair $(Z,\alpha) $ is said to be a generalised solution of the Hamilton-Jacobi problem, and when $\alpha$ is closed we say that $(Z,\alpha) $  is a standard solution the Hamilton-Jacobi problem. Note that in these conditions if a curve $\gamma$ in $Q$ is an integral curve of $Z$,
then the curve  $\alpha\circ \gamma$  in $T^*Q$ is    an integral curve of $X_H$.  This shows that the vector field $X_H$ is tangent to the image of $\alpha$. Moreover,   as a consequence, as $X_H$ is a vector field in $T^*Q$ and $T\pi_Q\circ T\alpha={\rm id}_{TQ}$, we have that $T\pi_Q\circ X_H\circ \alpha= T\pi_Q\circ T\alpha\circ Z=Z$, i.e. the vector field restriction of $X_H$ on  the image of $\alpha$ 
is projectable on $Z$. In this case  the vector field $Z$ in the pair $(Z,\alpha) $ can be derived from $\alpha$.

Observe that for each vector field $Y\in\mathfrak{X}(Q)$, if we recall that $\alpha^*\omega=d\alpha$, the 1-form $i(Z)d\alpha $ is such that 
$$(i(Z)d\alpha)(Y)=d\alpha (Z,Y)  =-\alpha^*\omega(Z,Y)= -\omega(T\alpha\circ Z, T\alpha\circ Y)\circ \alpha,
$$
while 
$$(\alpha^*(dH))(Y)=\alpha^*(i(X_H)\omega)(Y)=(i(T\alpha\circ Y)i(X_H)\omega)\circ \alpha=\omega(X_H\circ \alpha,T\alpha\circ Y)\circ \alpha,
$$
and consequently the relatedness condition 
$X_H\circ \alpha=T\alpha\circ Z$ is equivalent to 
\begin{equation}
i(Z)d\alpha+d(\alpha^*H)=0.\label{ZalfaH}
\end{equation}

 Remark that for a standard solution as the  1-form $\alpha$ is closed,  $d\alpha=0$, there exists a locally defined function $S$  in $Q$ such that $\alpha=dS$ and  the relatedness condition (\ref{ZalfaH})  can be replaced by $d(\alpha^*H)=0$, i.e $\alpha^*H$ is locally constant, or more explicitly 
 $$(dS)^*H=E\Longleftrightarrow H\left(q^1,\ldots,q^n,\pd S{q^i},\ldots,\pd S{q^i} \right)=E,
 $$
 which is the usually known as Hamilton-Jacobi equation. Once a solution of this equation has been obtained we can define the vector field $X^S \in\mathfrak{X}(Q)$ by $X^S=T\pi\circ X_H\circ dS$,
 and then $dS(Q)$ is an invariant under $X_H$ submanifold such that for each integral curve $\gamma$ of $X^S$, the curve $dS\circ \gamma$ is  an integral curve of $X_H$,  and in this sense the vector field $X^S$ in $Q$  is a reduction of $X_H$  which describes, at least partially,  the dynamics defined by $X_H$ in $T^*Q$. In this sense the integral curves of $X_H$ are completely in one of such invariant  submanifolds. 
 
 In order to fully solve the problem we need an appropriate $n$-parameter family of functions $S_{\lambda}$, with $\lambda=(\lambda_1,\ldots,\lambda_n)$, $\lambda_i\in \mathbb{R}$, what is usually known as a complete solution of the  Hamilton-Jacobi equation. More details and applications can be found in \cite{NRRM9}.
 
 \section{Conclusions}
 
 The theory of  integrability of systems  of  differential equations has been analysed from a geometric perspective through the properties of their associated vector fields. 
 The reduction process consists on finding a more easily integrable  related system whose solutions give us an at least partial information on the solutions of the original system. 
 The search for such simpler related systems is based on the use of first integrals and infinitesimal symmetries of the vector field or, more generally, invariant tensor fields.
  So, in Section 2 we have revisited the well-known classical Lie theorem and a more recent generalised result, without recourse to additional compatible structures, 
  while in Section 3 the r\^ole of invariant
 tensor fields in integrability has been discussed with a special emphasis on invariant (1,1)-tensor fields leading to symmetry generators and Lax equations with their correspondig 
 first integrals. In Section 4, after a brief summary  of the geometric Hamiltonian  and Lagrangian mechanics as well as the explicit formulation of Noether theorem in both frameworks,
  the Arnold-Liouville integrability has been  described. The meaning of non-Noether symmetries and how to find such symmetries have been discussed in Section 5.
  The usefulness  of invariant volume forms and its relation to Jacobi multipliers on oriented manifolds has been displayed in Section 6 as a prelude to the geometric approach to
   Hojman symmetry  briefly presented in Section 7. Finally, the geometric approach to Hamilton-Jacobi equation has also been analysed from the perspective of the search of 
   related systems.  according to the approach to reduction and integrability developed in this paper.
 
 \section*{Acknowledgements}

  Financial support from the projects of  Spanish Ministerio de Ciencia, 
Innovaci\'on y Universidades 
PGC2018-098265-B-C31  and Gobierno de Arag\'on $48\_20$R  An\'alisis y F\'{\i}sica Matem\'atica, is acknowledged.

\end{document}